\documentclass[runningheads]{llncs}
\usepackage{hyperref}

\usepackage[normalem]{ulem}
\usepackage{multicol}
\usepackage{comment}
\usepackage[pdftex]{color, graphicx}
\usepackage{subfig}
\usepackage{amsmath}
\usepackage{cleveref}

\usepackage{amssymb}
\usepackage{url}
\usepackage{xcolor}
\usepackage{float}
\usepackage{multirow}
\usepackage{algorithm}
\usepackage{algorithmic}
\usepackage{verbatim}
\usepackage{booktabs}
\usepackage{fancyhdr}
\title{Which bills are lobbied? Predicting and interpreting lobbying activity in the US\thanks{We thank the Center for Responsive Politics (OpenSecrets.org) for making their lobbying data available.}}

\author{Ivan Slobozhan\inst{1} \and
Peter Ormosi \inst{2} \and
Rajesh Sharma\inst{1}}
\authorrunning{I. Slobozhan et al.}
\institute{Institute of Computer Science, University of Tartu, Estonia, \email{ivan.slobozhan@gmail.com, rajesh.sharma@ut.ee, }
\and Centre for Competition Policy, University of East Anglia, UK \newline \email{p.ormosi@uea.ac.uk}}
\begin{document}
\maketitle

\begin{abstract}
%This paper looks at the application of text classification methods to help identify legislation that has been targeted by lobbying activities. 
Using lobbying data from OpenSecrets.org, we offer several experiments applying machine learning techniques to predict if a piece of legislation (US bill) has been subjected to lobbying activities or not. We also investigate the influence of the intensity of the lobbying activity on how discernible a lobbied bill is from one that was not subject to lobbying. We compare the performance of a number of different models (logistic regression, random forest, CNN and LSTM) and text embedding representations (BOW, TF-IDF, GloVe, Law2Vec). We report results of above 0.85\% ROC AUC scores, and 78\% accuracy. Model performance significantly improves (95\% ROC AUC, and 88\% accuracy) when bills with higher lobbying intensity are looked at. We also propose a method that could be used for unlabelled data. Through this we show that there is a considerably large number of previously unlabelled US bills where our predictions suggest that some lobbying activity took place. We believe our method could potentially contribute to the enforcement of the US Lobbying Disclosure Act (LDA) by indicating the bills that were likely to have been affected by lobbying but were not filed as such. 


\end{abstract}

%\textbf{Keywords:} missing data, multilayer networks, multiplex.

\section{Introduction}\label{sec:Intro}
\vspace{-3pt}
Lobbying consumes a significant amount of resources, which surpasses the money spent for example on campaign contributions. Opensecrets.org reports that lobbying expenditure reached around \$3.55 billion in 2010 (although it has started declining slowly since then, dropping to \$3.24 billion by 2013). US lobbying regulations ensure that much of the lobbying activities are disclosed to the public. As a result, there is ample information on the particulars of lobbying activities, and the access to this large amount of data has spurred numerous empirical works on lobbying.\footnote{For an overview see: \cite{de2014advancing}.}

The main contribution of this paper is a novel way to gauge whether a piece of legislation was lobbied or not. For this, we start on the premise that lobbying changes the text of legislation in a way that makes them discernible from non-lobbied legislation. Take \textit{rent seeking} for example. When businesses compete they earn normal profit as a result of the competitive process in the market. To increase profits, businesses will have to either collude, or monopolise the market, both of which would be blocked by antitrust agencies. The easiest way for companies to achieve super-normal profit is by lobbying governments to introduce laws and regulations that ensure that they are sheltered from competition. The economics literature calls this phenomenon rent seeking, referring to the objective of lobbying businesses to appropriate this \textit{rent} (i.e. super-normal profit). Rent seeking is hugely harmful for society, firstly because large amounts of resources are spent on a non-productive activity (lobbying), but also because the resulting markets are less competitive, meaning higher prices and therefore reduced welfare for consumers. We posit that if these legislative provisions, offering preferential treatment to certain interest groups, are similar across the various pieces of legislation, then the text of lobbied legislation should be discernible from non-lobbied ones. 

For this we rely on a database of lobbying activity in the US, and experiment with a number of text classification methods. In this respect our work diverges from previous works that apply text classification to expedite and improve the handling of large amounts of legal documents. By training a model to distinguish between lobbied and non-lobbied bills, our main objective is to improve legal analysis by discovering classification rules that had been unknown to human analysts.

This is important for multiple reasons. First of all, records on whether a bill had been lobbied may be incomplete. A classification algorithm could help ascertain if unlabelled bills have been lobbied or not. Second, although the US system is more transparent, the same is not true in jurisdictions where lobbying regulations are relatively new. For example, in the European Union there is very little information on the laws that are targeted by lobbyists. Using a model trained on US law we could investigate the use of transfer learning together with a much smaller sample of hand-labelled EU data to work on a model fitted to EU laws. Finally, our fitted model can also be informative for gauging the amount of rent seeking in the economy. Although not all lobbying activities should be considered as rent seeking, lobbying facilitates rent seeking.\footnote{In a similar logic as in \cite{laband1988social}} Moreover, \cite{lopez1994rent} estimated that lobbying activity accounts for around 2/3 of all rent seeking related welfare loss, with the figure being higher in more concentrated, and lower in less concentrated industries. 

As another contribution, the paper also tests the impact of more intensive lobbying. From the economics and finance literature we know that stakeholders with the largest expected profits from favourable policies and regulations are most likely to lobby most intensively\cite{hill2013determinants}. For this reason we expected more intensive lobbying associated with more discernible (for the algorithm) features when compared to non-lobbied legislation.

Using standard natural language processing (NLP) tools, we train a number of different models to classify bills into lobbied and non-lobbied groups. In particular, we used logistic regression, random forest and neural networks models. We achieve above 0.85 AUC and accuracy of 78\%. Moreover, we show that lobbying intensity improves model performance, up to 0.95 AUC and 88\% of accuracy implying that intensively lobbied bills are more different from non-lobbied ones (following our assumption that these are more likely to be subject to rent-seeking).

The rest of the paper is organised as follows. The next Section describes the literature review in this domain. In Section \ref{sec:data}, we introduce the dataset, and Section \ref{sec:evl} describes the results of our analysis. Finally, we conclude in Section \ref{sec:conl} with some future directions.
\section{Related works}\label{sec:rel}
\vspace{-3pt}
% In this section, we offer a brief overview of three streams of literature related to analysis of lobbying data: exploratory analysis (Section \ref{sec:relWork:Exp}), Predictive analysis (\ref{sec:relWork:Pred}), and network analysis (\ref{sec:relWork:Pred}).

In general, there is an increasing amount of literature that applies NLP in the legal domain.\footnote{\cite{dale2019law} gives an overview of the relevant literature.} Some of these focus more on solutions to automate summarising legal texts, such as court rulings \cite{farzindar2004legal} or \cite{hachey2006extractive}, applying SVM and naive Bayes classification of individual sentences to Bag of Words, TF-IDF, and dense features in order to improve summary precision. 

A subset of these applied NLP works in law draws on text classification methods. For example, \cite{boella2011using} use text classification methods (TF-IDF for feature extraction and SVM for text classification) in order to classify which domain a legal text belongs to. In another paper, \cite{li2018law} propose a semi-supervised learning method to classify legal texts. In this model the first step is the unsupervised learning of text region embedding, which is then fed into a supervised CNN. 

Finally, a large number of NLP applications in law focus on prediction.  \cite{wongchaisuwat2017predicting} set out to predict various aspects of patent litigation, with mixed results. Other works focus on the prediction of court rulings, such as the European Court of Human Rights (ECRH) decisions by \cite{aletras2016predicting}, or French Supreme Court rulings by \cite{sulea2017predicting}. 

There is a well-established body of literature on lobbying, and it is beyond the remits of this paper to provide an overview of these. In a systematic review of the relevant empirical works, \cite{de2014advancing} takes account of the main strands of empirical papers and the challenges to empirical research on lobbying. It also discusses the advantages, disadvantages, and effective use of the main types of data available. Nevertheless, none of these reviewed works used methods similar to ours.

The closest we can relate our paper to previous literature is in the area looking at the impact of lobbying on the specific bills they are targeting. \cite{grasse2011influence} found a direct association between lobbying activities and bill outcomes, and that public attention reduces the effects of lobbying efforts, suggesting that lobbying is most effective when focused on less salient issues. In another paper, \cite{you2017ex} looks at the difference between bills that were lobbied ex post and those lobbied before they were passed. Finally, in \cite{grossmann2013lobbying} the authors look at the determinants of interest group lobbying on particular bills after the bills have been passed, and identifies the areas where lobbying focusing on the implementation (rather than the formation) of legislation is more likely.

%
\section{Dataset}\label{sec:data}
\vspace{-3pt}
The data was downloaded from the Center for Responsive Politics.\footnote{\url{https://www.opensecrets.org}} The dataset contains detailed information on a large number of lobbying instances. For the purposes of this paper our focus is on data on the bills that were lobbied. At the time of downloading the data (October 2018) the data contained information on lobbying activities related to 54,713 US bills. Table \ref{tab:types} shows the breakdown of these bills by bill type - most of them are House of Representative Bills or Senate Bills.

\vspace{-14pt}
\begin{table}
\centering
    \caption{Lobbied bills by bill type}
\begin{tabular}{lr}
\toprule
bill\_type   &  n      \\
\midrule
H.Con.Res.  &    334 \\
H.J.Res.    &    346 \\
H.R.        &  31862 \\
H.Res.      &   1290 \\
S.          &  19915 \\
S.Con.Res.  &    150 \\
S.J.Res.    &    177 \\
S.Res.      &    597 \\
\bottomrule
\end{tabular}
    \label{tab:types}
%\vspace{-5pt}
\end{table}

We downloaded all bills available in text format from the US Congress' website.\footnote{An example of a House Bill is given here: \url{https://www.congress.gov/bill/114th-congress/house-bill/3791/text}.} We then marked out the bills that had been lobbied, and then matched it with a similar sample ($n=48411$) of other bills where we had no evidence that there was any lobbying and thus, we assumed that there was no lobbying in these cases.\footnote{This is a realistic assumption. In the US, lobbying activities (above a certain threshold) need to be disclosed, and non-compliance can result in a pecuniary sanction (fine) or, in some cases up to 5 years imprisonment. Nevertheless, later in our paper we revisit this assumption.} This resulted in a total sample of 103,243 labelled bills (54,377 lobbied, 48,530 non-lobbied). Table \ref{tab:subjects} shows the breakdown of the sample into subject areas.

%\vspace{-8pt}
\begin{table}[!h]
\centering
\caption{Number of bills by subject area and lobbying activity}
\scalebox{0.75}{
\begin{tabular}{lrr}
\toprule
subject  &   not lobbied &   lobbied \\
                                           &       &       \\
\midrule
        Agriculture and Food               &   675 &  1130 \\
        Animals                            &   206 &   322 \\
        Armed Forces and National Security\textbackslash ... &  3001 &  4067 \\
        Arts, Culture, Religion            &   304 &    58 \\
        Civil Rights and Liberties, Minorit... &   507 &   382 \\
        Commemorations                     &  3934 &   414 \\
        Commerce                           &   756 &  1411 \\
        Congress                           &  3928 &   849 \\
        Crime and Law Enforcement          &  1949 &  2622 \\
        Economics and Public Finance       &   716 &   975 \\
        Education                          &  1824 &  2474 \\
        Emergency Management               &   546 &   799 \\
        Energy                             &   716 &  1847 \\
        Environmental Protection           &   692 &  1452 \\
        Families                           &   370 &   259 \\
        Finance and Financial Sector       &   723 &  2086 \\
        Foreign Trade and International Fin... &  3657 &  3567 \\
        Government Operations and Politics\textbackslash ... &  2719 &  2664 \\
        Health                             &  3364 &  6943 \\
        Housing and Community Development\textbackslash r... &   405 &   806 \\
        Immigration                        &   836 &  1245 \\
        International Affairs              &  4107 &  2008 \\
        Labor and Employment               &   786 &  1355 \\
        Law                                &   558 &   673 \\
        Native Americans                   &   549 &   653 \\
        Private Legislation                &   838 &   203 \\
        Public Lands and Natural Resources\textbackslash ... &  2728 &  2883 \\
        Science, Technology, Communications... &   595 &  1205 \\
        Social Sciences and History        &    64 &    18 \\
        Social Welfare                     &   726 &   771 \\
        Sports and Recreation              &   420 &    93 \\
        Taxation                           &  3485 &  5679 \\
        Transportation and Public Works... &  1120 &  2114 \\
        Water Resources Development        &   607 &   644 \\
\bottomrule
\end{tabular}
}
\label{tab:subjects}
\end{table}
\vspace{-3pt}

\begin{figure}[!htp]
    \centering
    \includegraphics[width=10cm]{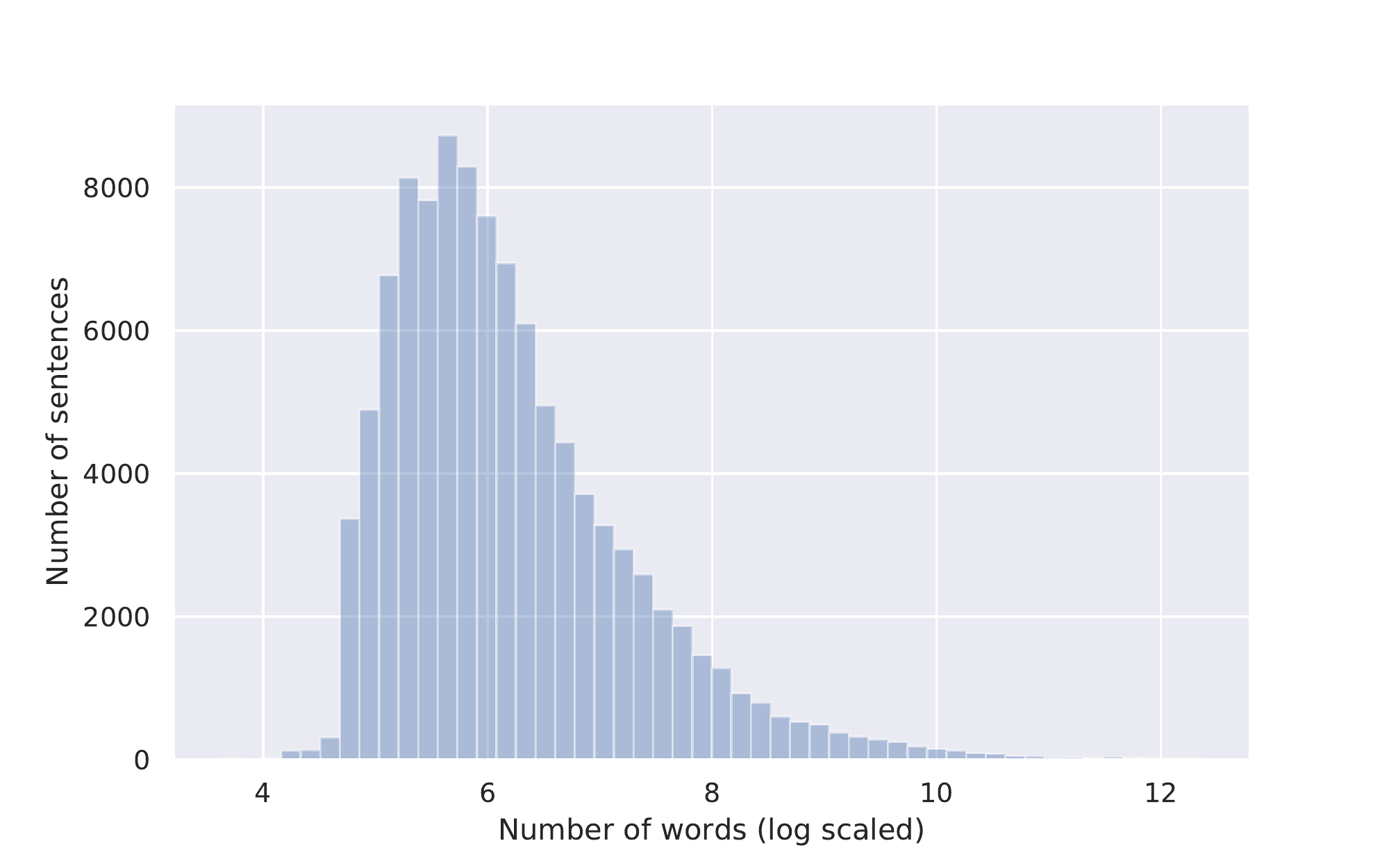}
    \caption{The distribution of the number of the words}
    \label{fig:distribution}
\end{figure}

One of the challenges that we had in our experiments is the size of the bills. Figure \ref{fig:distribution}  shows the distribution of the number of words in the documents. The average bill length was 16,181 words, but the bill length distribution was highly skewed, some bills with over a million words. For computational ease we dropped all bills that were longer than 150,000 words, leaving us with a sample of 101,755 labelled bills. 

Finally, we tested how much lobbying-intensity affected classification performance. The reason we thought this was important, was that lobbying activities are largely heterogeneous. For example, some lobbying activities might not lead to changes in the text of the legislation. Intuitively, less intensive lobbying is less likely to lead to any changes in legislative provisions. Also, some lobbying can be benign, and more likely to make only small changes to a given piece of legislation. On the other hand, for lobbying driven by rent seeking the same is probably not true. We posit that businesses with more to gain from lobbying (rent seeking) are more likely to lobby intensively, and therefore lobbying intensity is more likely to be correlated with having provisions in a bill that make these lobbied bills different from non-lobbied ones. To test this, we introduce the information we had on lobbying intensity into the way we labelled our data.

In Table \ref{tab:intensity} we show the number of bills associated with different levels of lobbying intensity. Around a half of the bills were not lobbied, roughly another quarter of them were lobbied between 1-10 times, and the rest even more frequently.

For our analysis we created different labels to reflect lobbying intensity. Let $lobbied$ denotes the number of times a bill was lobbied, then, we created three versions of the datasets using following logic: 
\begin{equation}
D_{1} =\begin{cases}1 & \text{if } lobbied \geq 1\\0 & \text{if } lobbied = 0\end{cases}
\end{equation}

\begin{equation}
D_{2} =\begin{cases}1 & \text{if } lobbied  \geq 10\\0 & \text{if } lobbied= 0\end{cases}
\end{equation}

\begin{equation}
D_{3} =\begin{cases}1 & \text{if } lobbied \geq 50\\0 & \text{if } lobbied= 0\end{cases}
\end{equation}

\vspace{-4pt}
\begin{table}[]
\centering
\caption{Number of bills exposed to different levels of lobbying intensity}
\begin{tabular}{p{4cm}c}
\toprule
Number of times lobbied & Number of bills \\
\midrule
(0.0]				 &	  48530 \\
(1.0, 5.0]           &    18511 \\
(5.0, 10.0]          &     7338 \\
(10.0, 50.0]         &    14924 \\
(50.0, 100.0]        &     5072 \\
(100.0, 200.0]       &     3836 \\
(200.0, 500.0]       &     3003 \\
(500.0, 1000.0]      &     1136 \\
(1000.0, 10000000.0] &      893 \\
\bottomrule
\end{tabular}
\label{tab:intensity}
\end{table}
\vspace{-1pt}
%\clearpage

We used these labels to create three balanced 'datasets', with $D_1$ mapping out dataset 1 and so on. The respective sample sizes of datasets for label $D_{1}$, $D_{2}$ and $D_{3}$, are 103,243, 59,834, and 28,132 bills (13,217 lobbied and 14,915 non-lobbied).   %
\section{Evaluation}\label{sec:evl}
\vspace{-3pt}
In this section, we present the results of our evaluation. First, we describe the algorithms (Section 4.1), next, the metrics we used for evaluating our approach (Section 4.2), then, the overall approach for text pre-processing, feature generation, and hyperparameter tuning is discussed (Section 4.3) and finally, we present the results (Section 4.3).

\subsection{Problem modeling and Algorithms}
\vspace{-2pt}
We modeled the problem as a binary classification task. Our objective was to classify a given document into one of the two categories, that is, if the document has been lobbied or not. To solve this task, we used four types of algorithms: logistic regression, random forests, \cite{RF-Breiman2001}, and neural networks, more specifically, using recurrent neural networks (LSTM) and Convolution neural networks (CNN) for text classification. We also experiment with various feature extraction algorithms such as bag of words (BOW), term frequency-inverse document frequency (TF-IDF), word embeddings for neural networks, and a domain specific Law2Vec embeddings, which we chose, given our task relates to legal documents.\cite{chalkidis2019deep}

\subsection{Metrics}
\vspace{-2pt}
We checked the performance of our three algorithms using two main classification metrics: accuracy (ACC) and area under a receiver operating characteristic curve (AUC ROC). 
\begin{enumerate}
    
\item  \textbf{Accuracy: }is defined as a ratio of correctly classified observations to the number of all observations. The perfect binary classifier will have 100\% accuracy, and random binary classifier has 50\% of accuracy on a balanced dataset.

\item \textbf{AUC ROC: }is equal to the probability that a classifier will rank a randomly chosen positive observation higher than a randomly chosen negative one. AUC ROC is calculated by plotting true positive rate against the false-positive rate at different thresholds. True positive rate is the proportion of actual positives that are identified correctly, and the false-positive rate is the ratio between the false positives and the total actual negative cases. After that the area of this curve is calculated to get AUC ROC. The perfect binary classifier will have AUC ROC equal to 1, and in a random binary classifier ROC AUC equals to 0.5.
\end{enumerate}

\subsection{Approach}
\vspace{-2pt}
Our pipeline consists of the following three steps. 
 
 \begin{enumerate}

 \item \textbf{Data Cleaning: } We applied conventional text pre-processing steps to our raw documents (the text of bills). In particular, we lowercase the text, deleted numbers, english stopwords, law stopwords, special characters and punctuation from the text. After that, for each word, we perform lemmatization. For the logistic and random forest, we did not truncate or pad the sentences. However, due to computational issues in using the LSTM and CNN models, we set a max size for the sentence to be 2500 words after the first part of our pre-processing pipeline. After that, we truncated all the sentences which were above the length 2500. Those sentences that are below the defined length are padded with special tokens.
 %\fixme{@Ivan: I could not understand. Check English please.}
 \item \textbf{Feature creation: }Next, we transformed the preprocessed text into a set of features that can be fitted into a machine learning model. For logistic regression and random forest we used TF-IDF on bag of n-grams and bag of words approach for text representation with a dictionary size set to 25000. For the neural network, we used 300 dimensional GloVe word embeddings,\cite{Pennington14glove} and Law2Vec 200 dimensional embeddings.

 \item \textbf{Hyperparameter tuning: }Finally, we hypertuned the model using algorithm-specific parameters on the validation dataset, where we tried to find parameters that maximize AUC ROC. For example, we searched for the best value of $n$ for the models trained using TF-IDF on a bag of n-grams. In particular, for logistic regression, we grid searched the best parameters for regularization strength, and penalty type. In case of random forest, we experimented with a larger set of hyperparameters such as maximum tree depth, splitting criteria, minimum samples per split, and minimum leaf size. For neural networks, we experimented only with different optimization algorithms and the number of recurrent layers. 
\end{enumerate}

After the best parameters have been determined, we then used the validation dataset to find the threshold that maximizes the accuracy. Finally, we ran our experiments on a test set and reported the accuracy using the best threshold we found on the validation dataset. 

\subsection{Results}
%\vspace{-2pt}
We perform our experiments using three different versions of the dataset, which aims to capture lobbying intensity through different labeling, as explained in Section \ref{sec:data}. We denoted these as Labelling 1, 2, and 3. The corresponding results are reported in Table \ref{Tbl:label3}. In each of these three versions of the dataset, we split the data into train, validation, and test sets with proportions of 72\% for train, 8\% for validation, and 20\% for test, respectively.

%\vspace{-1pt}

% We name these datasets as dataset 1 and dataset 2 and are characterized based on the number of times it has been lobbied.
% \begin{enumerate}

% \item  \textbf{Dataset 1} is a dataset with the following ground truth: 0 if the number of lobbying is equal to 0 and ground truth is 1 if lobbying is larger than 1. 

% \item  \textbf{Dataset 2} is a dataset with the following ground truth: 0 if the number of lobbying is equal to 0 and ground truth is 1 if lobbying is larger than 10. 

% \item \textbf{Dataset 3} is a dataset with the following ground truth: 0 if lobbying is equals to 0 and 1 if lobbing is larger than 50. 
% \end{enumerate}
% \usepackage{multirow}

%\vspace{-10pt}
% Please add the following required packages to your document preamble:

%\vspace{-10pt}

% Please add the following required packages to your document preamble:
% \usepackage{multirow}

It appears that the best performing model and feature sets are logistic regressions with TF-IDF, and LSTM with GloVe embeddings. In many text classification applications neural networks with word embeddings work better than other models \cite{goldberg2017neural}, especially when researchers have access to a large corpus of text data. We observe this in our case as well, especially if one looks at Table \ref{Tbl:label3}. Regarding the word embedding representations, we TF-IDF provides the best AUC ROC and accuracy on the test sample. Interestingly, Law2Vec is slightly outperformed by both GloVe and TF-IDF.

The performance of the logistic model stands out. This would suggest that in our binary classification problem the classes (lobbied - non-lobbied) are linearly separable. On the other hand, the performance of deep learning models did not exceed the logistic regression, which is likely to be down to the relatively small size of our sample. In the most informative model that compares high-intensity lobbied bills with non-lobbied bills (Labelling 3), we have 13,217 lobbied and 14,915 non-lobbied bills. This is small, especially given that our median bill length is 4790 words (the mean is 10413), so each observation contains a very large number of features. It is likely that there are complex non-linear relationships between these features and the classes, but to fully explore this complexity we would need much larger samples. Another possible reason is that the logistic regression model, compared to neural network do not need sophisticated and time consuming hyperparameter tuning. Due to computation limitations we are not able to explore a large set of possible hyperparameters for the neural network models. On the other hand, it is true that the good performance of the logistic model also suggests that there are clear features (such as the frequency of specific n-grams) in these bills, which form a linear relationship with our two classes. This is crucial in our application (text classification in Law), where interpretability is very important for users. In Section 4.5 we provide an introduction to these key features.

%\subsubsection{Which features are important in predicting the results}
\begin{table}[]
\caption{Classification results - results for our three labels}
\label{Tbl:label3} 
\resizebox{\textwidth}{!}{%
\begin{tabular}{|l|c|c|c|c|}
\hline
\multicolumn{1}{|c|}{\multirow{2}{*}{Model}} & \multicolumn{2}{c|}{Validation} & \multicolumn{2}{c|}{Test} \\ \cline{2-5} 
\multicolumn{1}{|c|}{}          & AUC ROC & ACC.    & AUC ROC & ACC.    \\

\hline
\multicolumn{5}{l}{} \\ 
\multicolumn{5}{l}{Labelling 1} \\ 
\hline
Logistic regression (TF-IDF)    & 0.8566  & 77.51\% & 0.8609  & 78.19\% \\ \hline
Logistic regression (BOW)       & 0.8233  & 74.58\% & 0.8253  & 74.72\% \\ \hline
Random forest (TF-IDF)          & 0.8451  & 76.23\% & 0.8498  & 76.72\% \\ \hline
LSTM (GloVe 300d. embeddings)   & 0.8658  & 78.12\% & 0.8652  & 78.31\% \\ \hline
LSTM (Law2Vec 200d. embeddings) & 0.8514  & 77.24\% & 0.8503  & 77.21\%  \\ \hline
CNN (GloVe 300d. embeddings)    & 0.8520  & 77.14\% & 0.8550  & 77.68 \% \\ \hline
CNN (Law2Vec 200d. embeddings)  & 0.8529  & 76.71\% & 0.8501  & 76.71\% \\ 
\hline
\multicolumn{5}{l}{} \\ 
\multicolumn{5}{l}{Labelling 2 } \\ 
\hline
Logistic regression (TF-IDF)    & 0.9318  & 85.95\% & 0.9321  & 85.73\% \\ \hline
Logistic regression (BOW)       & 0.8337  & 82.20\% & 0.8920  & 81.19\% \\ \hline
Random forest (TF-IDF)          & 0.9169  & 83.83\% & 0.9179  & 83.39\% \\ \hline
LSTM (GloVe 300d. embeddings)   & 0.9334  & 86.14\% & 0.9300  & 85.61\% \\ \hline
LSTM (Law2Vec 200d. embeddings) & 0.9204  & 84.35\% & 0.9222  & 84.25\% \\ \hline
CNN (GloVe 300d. embeddings)    & 0.9251  & 84.95\% & 0.9280  & 85.16\% \\ \hline
CNN (Law2Vec 200d. embeddings)  & 0.9240  & 84.98\% & 0.9257  & 84.87\% \\

\hline
\multicolumn{5}{l}{} \\ 
\multicolumn{5}{l}{Labelling 3 } \\ 
\hline
Logistic regression (TF-IDF)    & 0.9557  & 88.79\% & 0.9548  & 88.79\% \\ \hline
Logistic regression (BOW)       & 0.9129  & 84.54\% & 0.9128  & 84.16\% \\ \hline
Random forest (TF-IDF)          & 0.9431  & 86.69\% & 0.9430  & 85.80\% \\ \hline
LSTM (GloVe 300d. embeddings)   & 0.9505  & 89.38\% & 0.9447  & 87.86\% \\ \hline
LSTM (Law2Vec 200d. embeddings) & 0.9406  & 86.91\% & 0.9393  & 86.37\%  \\ \hline
CNN (GloVe 300d. embeddings)    & 0.9519  & 88.70\% & 0.9487  & 88.00\% \\ \hline
CNN (Law2Vec 200d. embeddings)  & 0.9450  & 87.14\% & 0.9459  & 87.05\%  \\ \hline
\end{tabular}%
}
%\vspace{-4pt}
\end{table}

Comparison of the three sets of results clearly indicates that the prediction improves as we re-define our label in terms of lobbying intensity. In the first experiment (top section of Table \ref{Tbl:label3}) we compare bills that were not lobbied, with bills that were lobbied, irrespective of the number of times. This provides the worst results. This is in line with intuition: it is likely that bills that were lobbied only once are not hugely different from those that were not lobbied at all (for example, the lobbying might have been for a benign, minor correction of the text, or the lobbying might have not successfully changed the text of the legislation at all).

For Labelling 2 (middle section of Table \ref{Tbl:label3}) and 3 (bottom section of Table \ref{Tbl:label3}) the results show improvement. In these experiments we compared bills that were not lobbied with bills that were lobbied intensively, at least 10, and at least 50 times respectively. The results suggest that the difference between the text of lobbied and non-lobbied bills becomes more discernible where there is more intensive lobbying. Put differently, a bill that was not lobbied is more similar to a bill that was lobbied only once, than to a bill that was lobbied, say, 20 times.

\subsection{Interpreting the classification results}
In this subsection, we make an attempt to explain which features played the most important role in generating our results. The good performance of the logistic model means that we have a better chance to interpret what features are driving our classification algorithm. 

\begin{comment}
For this reason we report the most important features (twenty words that had the most important negative (Figure \ref{fig:negative_words}), and positive (Figure \ref{fig:positive_words}) effects on the classification. These were extracted using the logistic regression model with features extracted using tf-idf algorithm on a bag of unigrams and bigrams, trained on the dataset with Labelling 3 (as this labelling gave us the best performance). 
\end{comment}

We extracted the most important features using the logistic regression model with tf-idf algorithm on a bag of unigrams and bigrams, trained on the dataset with Labelling 3 (as this labelling gave us the best performance). Among the most important features we can find \textit{congress appropriation}, which refers to appropriation bills – i.e. bills that decide on how to allocate federal funds to various specific federal government departments, agencies and programs. Increased lobbying activity of these bills that directly decide on how to spend money are not surprising.

Scanning through the 100 most important features,\footnote{A list of the top 100 positive and negative features is given in the Appendix.} one also finds a list of senator names: \textit{Cartwright}, \textit{Polis}, \textit{Roe}, \textit{Murphy}, \textit{Reed}, \textit{Kelly}. It is likely that bills introduced by these Senators received more lobbying than bills introduced by others, which is why we pick up their names among the top features.

The top feature list also contains a number of terms that are typically associated with legislation that limit competition in one way or another. Terms like \textit{exception}, \textit{reauthorization}, \textit{protection}, \textit{prevent}, \textit{copyright}, \textit{patent}, are possible signs of the regulatory protection of some market players, or the creation of regulatory monopolies through patents or copyrights.

Finally, one can also see patterns of the sectors and topics where more lobbying happens, such as finance: \textit{insurance}, \textit{health saving}, \textit{credit union}, \textit{share agreement}, \textit{flood insurance}, \textit{saving}, \textit{tax freedom}; public health: \textit{abortion}, \textit{care assistance}, \textit{overdose}, \textit{smoker}, \textit{cancer screening}; infrastructure: \textit{infrastructure}, \textit{building code}, \textit{federal land}; or associated with socially controversial topics: \textit{abortion}, \textit{marriage}, \textit{partnership}, \textit{ammunition}, \textit{gender identity}.

\subsubsection{Subject-level feature analysis}

\begin{figure}[!htp]
    \centering
    \includegraphics[width=10cm]{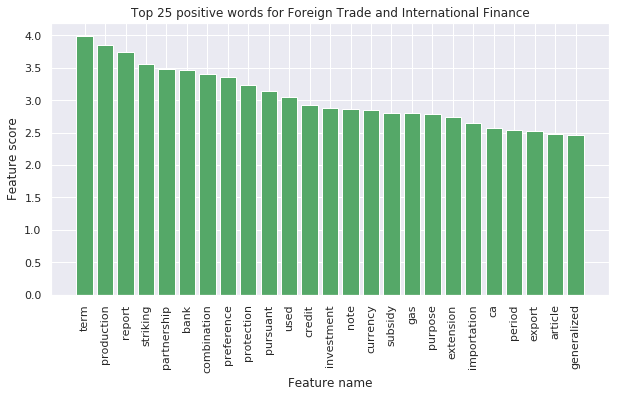}
    \caption{Most important positive words for Foreign Trade and International Finance}
    \label{fig:positive_words1}
\end{figure}

Looking at each subject more specifically can lead us to more fine-tuned feature importance discussions. Below we provide some examples. Looking at bills on Foreign Trade and International Finance, Figure \ref{fig:positive_words1} shows \textit{preference}, \textit{protection}, \textit{credit}, \textit{subsidy}, and \textit{extension} among the positive features (indicating higher probability of lobbying). This is not surprising, these terms are typically associated with various trade barriers, one of the prime manifestations of successful rent seeking lobbying by US-based producers. Other features, such as \textit{combination} and \textit{partnership} are signs of export/import partnership, which are often the subject of trade-related rent seeking activities. Of course, finding import, export, or currency (words that are inherent in trade related documents) among the important features shows that our selection of stopwords would have to be further fine-tuned to each subject area specifically.

\begin{figure}[!htp]
    \centering
    \includegraphics[width=10cm]{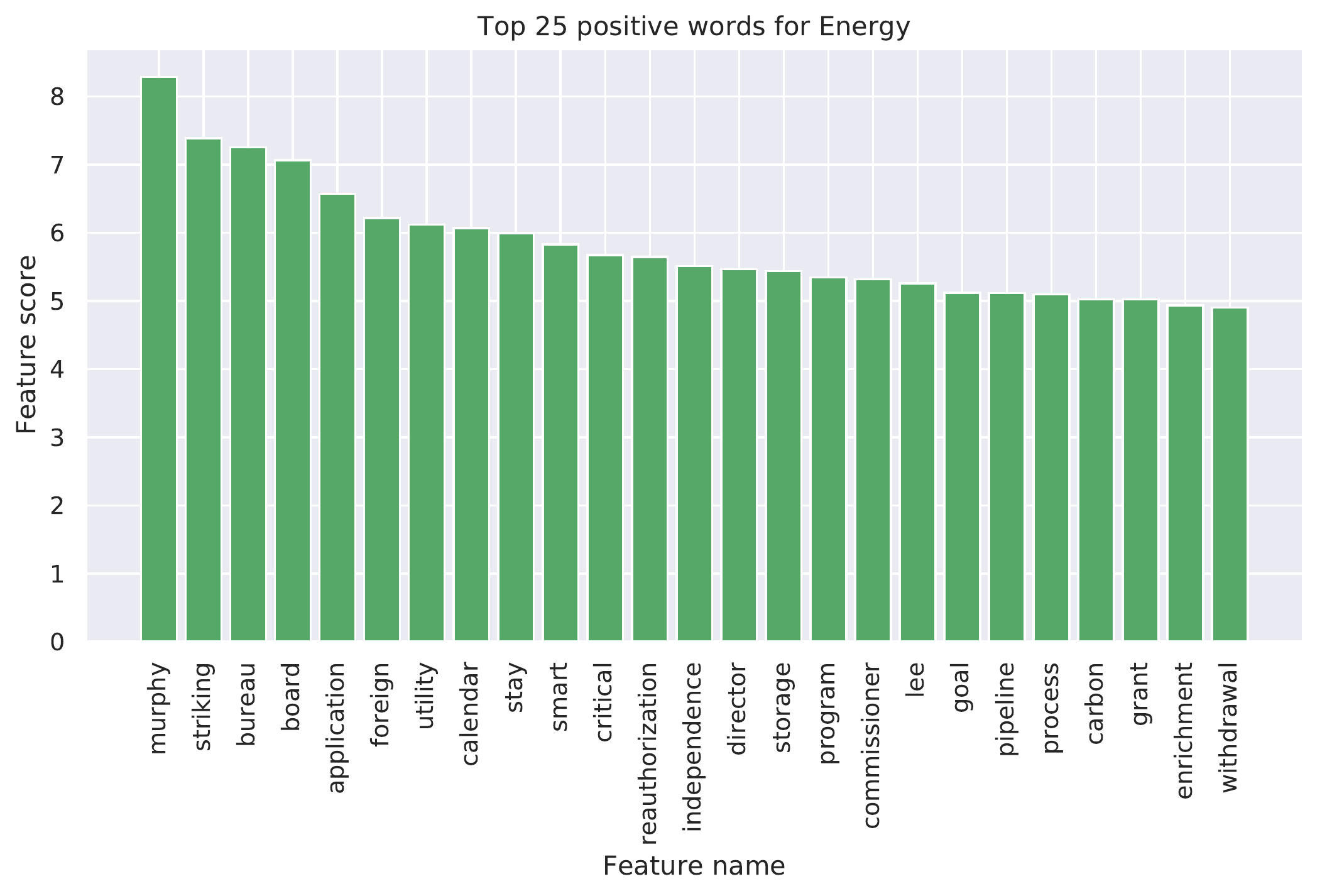}
    \caption{Most important positive words for Energy}
    \label{fig:positive_words2}
\end{figure}

The 25 most important features for the subject Energy, shown on Figure \ref{fig:positive_words2} also reveals a number of interesting patterns. For example, names, such as Murphy and Lee, probably names of senators, implying that bills introduced by them were more likely lobbied. The feature \textit{smart} is likely to refer to smart meters. Smart meters have a high penetration rate in the US (in 2018, there were over 86 million smart meters installed in the country\footnote{\url{https://www.eia.gov/tools/faqs/faq.php?id=108&t=3}}. There could be many reasons why these are subject to more intensive lobbying. Smart meter data can make energy distribution more efficient, but it also means the possibility of creating data monopolies, that are later very difficult to challenge to new entrants. There is also a view that smart meters can make consumer switching more difficult, which, again, would explain the increased interest to lobby relevant bills. \textit{Carbon} also appears among the more lobbied features. This is not surprising, given the prominent role of climate and carbon related lobbying.\footnote{See for example \cite{brulle2018climate}.}

\section{Application to unlabelled data}\label{sec:extl}
\vspace{-3pt}
As mentioned earlier, one of the limitations of using the OpenSecrects.org data is that it only labels the bills that were lobbied, making the implicit assumption that all unlabelled bills were not lobbied. As explained above, in the US, lobbying activities are required to be disclosed, violations of which can lead to severe penalties. Nevertheless, there has been over 14,000 such violations since 1995,\footnote{\url{https://www.hklaw.com/en/insights/publications/2017/11/what-is-the-lobbying-disclosure-act-lda}} which would suggest that non-compliance is a non-trivial problem. Our proposed approach below offers a way to verify if those bills that are not entered in the OpenSecrets.org database have been subject to similar lobbying activities as those that are listed by OpenSecrets.org. 

In our experiment, we downloaded all available bills from the US Congress' website (254,806 bills). As very old bills could have had different wordings, and short bills are likely to have limited amount of information for our analysis, we constrained this sample to bills after 1990, and bills that were at least 2000 word long, which left us with a sample of 81,998 bills. From the lobbied bills we only used the ones where there was intensive lobbying (50 instances or more), i.e. where we were most certain to find distinctive features due to the lobbying (13,217 bills). 

\begin{figure}[!htp]
    \centering
    \includegraphics[width=1\textwidth]{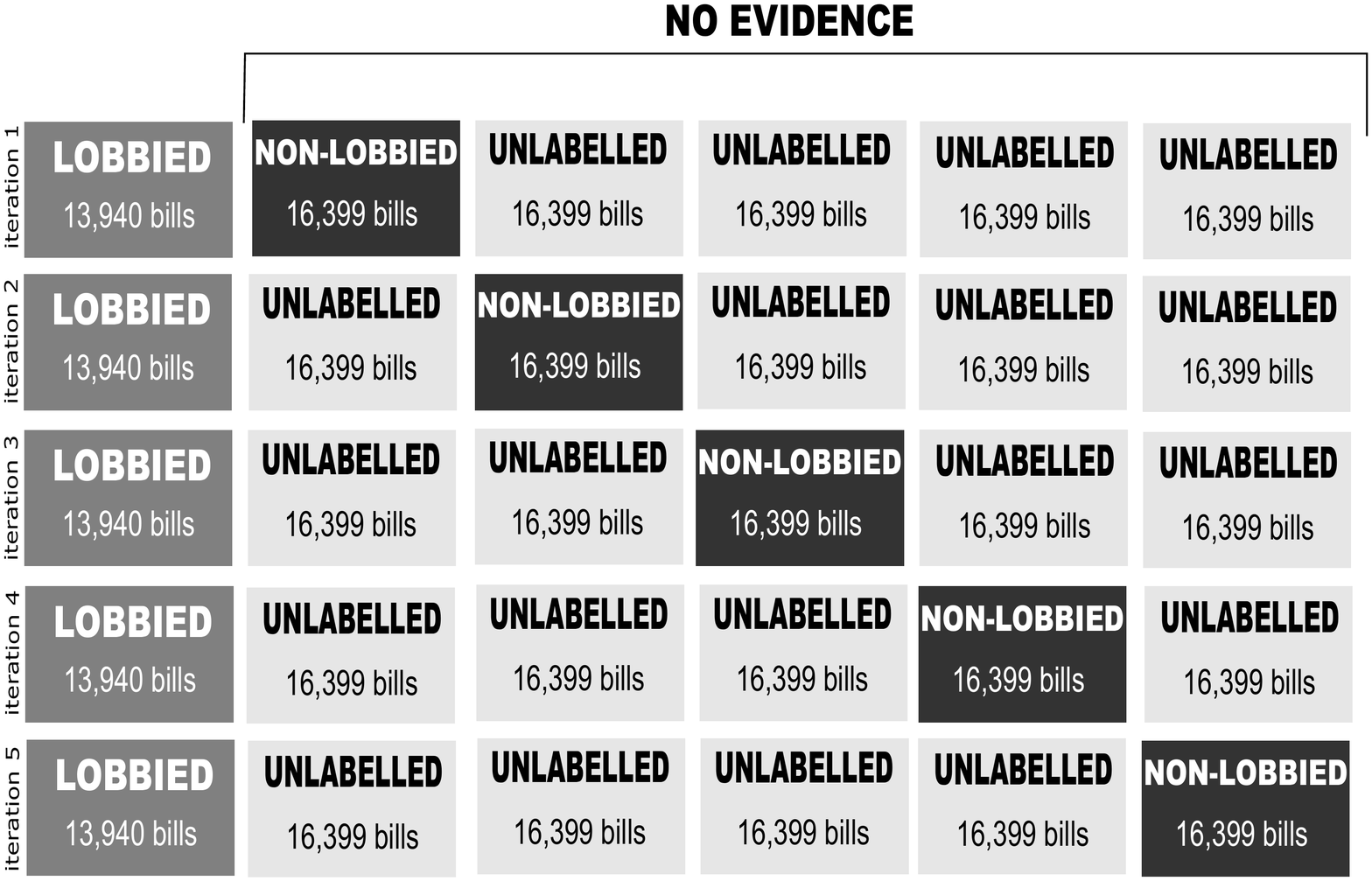}
    \caption{Extracting information from non-labelled data}
    \label{fig:iterations}
    \vspace{-6pt}
\end{figure}

First, to estimate a model that predicts lobbying in a bill, we took our 13,217 lobbied bills (labelled as lobbied), and used cross-validation to take 5 rotated samples (each consisting of 81,998/5=16,399 bills) from the unlabelled bills and labelled them as non-lobbied. This cross-validation exercise is shown on Figure \ref{fig:iterations}. Then we estimated our model (using a logistic model given its relatively good performance and speed) and deployed it on the remaining ‘unlabelled’ sample to predict the probability that a given unlabelled bill was lobbied. We then moved on to the next iteration, where we used the same lobbied sample, but another 16,399 unlabelled bills were selected and labelled as non-lobbied. Then we estimated our model for this new  set of labelled bills, and deployed it on the remaining sample, and so on. For each unlabelled bill and for each iteration, we stored the estimated probability that it was lobbied. The five batches in our iterations gave us 4 predictions for each unlabelled bill. We then took the average of these 4 predictions as a probability that an unlabelled bill was directly or indirectly affected by lobbying activity.

\begin{figure}[!htp]
    \centering
    \includegraphics[width=0.8\textwidth]{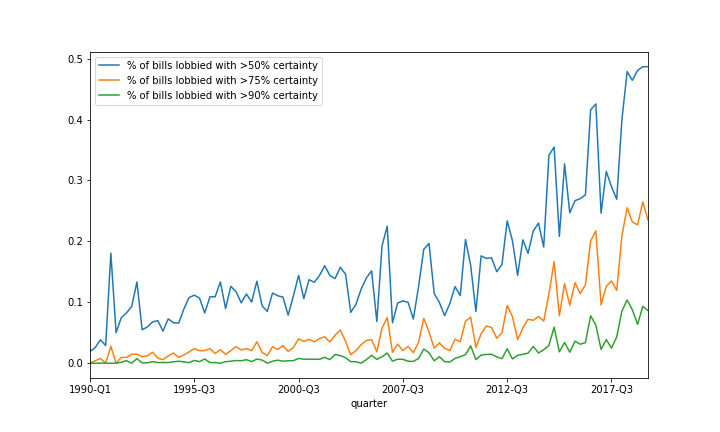}
    \caption{Proportion of non-labelled bills with evidence of lobbying}
    \label{fig:non_labelled_ts}
    \vspace{-6pt}
\end{figure}

Figure \ref{fig:non_labelled_ts} plots these average probabilities over time (calendar quarter of the release of the bill). This shows an increasing trend in the percentage of unlabelled bills being affected by lobbying, indicating, that for more recent bills, almost half had at least a 50\% probability that they were affected by lobbying (>50\% probability), and almost 10\% of unlabelled bills were predicted to have been lobbied with over 90\% probability. 

Finally, Table \ref{tab:unlabelled_subject} presents the proportion of unlabelled bills where we predicted a high probability of lobbying activity, broken down to subject areas. To preserve space we only report 10 subject areas with the highest probability. It shows that in subjects such as \textit{Energy}, \textit{Finance and Financial Sector}, \textit{Science and Technology}, and \textit{Health} around 5\% of the unlabelled bills were affected by lobbying with more than 90\% probability. When looking at the bills that were most likely have been lobbied, Table \ref{tab:topbills} lists the 25 bills which we estimated to be most likely to have been lobbied but were not recorded as such on OpenSecrets.org at the time of us accessing the data (Jan 2019). It is possible that there is a lag in recording lobbied bills, but even if this is the case, and if these bills were later added to the lobbied list, it would confirm that our model made the right prediction.

\begin{table}[]
\caption{Top 25 highest lobbying probability bills} \label{tab:topbills}
\scalebox{0.65}{
\begin{tabular}{lllll}
\toprule
bill\_name    & title                                                                                                                 & congress & \begin{tabular}[c]{@{}l@{}}average \\ probability\end{tabular} & subject                             \\
\midrule
S.2519        & GREENER Fuels Act                                                                                                     & 115th    & 0.9992                                                         & Energy                              \\
H.R.5212      & GREENER Fuels Act                                                                                                     & 115th    & 0.9992                                                         & Energy                              \\
H.Res.1082    & FAA Reauthorization Act of 2018                                                                                       & 115th    & 0.999                                                          & Congress                            \\
H.R.5473      & \begin{tabular}[c]{@{}l@{}}Better Pain Management Through Better \\ Data Act of 2018\end{tabular}                     & 115th    & 0.9981                                                         & Health                              \\
S.J.Res.63    & No title                                                                                                              & 115th    & 0.9974                                                         & Health                              \\
H.R.5189      & Opioid Abuse Crisis Act of 2016                                                                                       & 114th    & 0.9973                                                         & Health                              \\
H.R.104       & Leave Ethanol Volumes at Existing Levels Act                                                                          & 116th    & 0.9968                                                         & Environmental Protection            \\
H.R.6620      & \begin{tabular}[c]{@{}l@{}}Protecting Critical Infrastructure Against Drones \\ and Emerging Threats Act\end{tabular} & 115th    & 0.9966                                                         & Emergency Management                \\
S.Res.18      &                                                                                                                       & 114th    & 0.9962                                                         & Congress                            \\
S.2502        & \begin{tabular}[c]{@{}l@{}}Protecting Communities and Preserving the \\ Second Amendment Act of 2018\end{tabular}     & 115th    & 0.9961                                                         & Crime and Law Enforcement           \\
S.2202        & Renewable Fuel Standard Extension Act of 2007                                                                         & 110th    & 0.9959                                                         & Environmental Protection            \\
H.R.5785      & Jobs and Justice Act of 2018                                                                                          & 115th    & 0.9958                                                         & Social Welfare                      \\
H.R.5906      & \begin{tabular}[c]{@{}l@{}}Financial Services and General Government \\ Appropriations Act, 2019\end{tabular}         & 115th    & 0.9957                                                         & Economics and Public Finance        \\
H.R.4         & ARPA-E Act of 2018                                                                                                    & 115th    & 0.9957                                                         & Science, Technology, Communications \\
H.R.6258      & FAA Reauthorization Act of 2018                                                                                       & 115th    & 0.9957                                                         & Transportation and Public Works     \\
S.Res.18      & No title                                                                                                              & 113th    & 0.9955                                                         & Congress                            \\
H.R.6227      & DHS Reform and Improvement Act                                                                                        & 114th    & 0.9953                                                         & Emergency Management                \\
H.R.6381      & Grid Cybersecurity Research and Development Act                                                                       & 114th    & 0.9953                                                         & Energy                              \\
H.Con.Res.155 &                                                                                                                       & 114th    & 0.9952                                                         & Commerce                            \\
S.3523        & SAVING AMERICAN ENERGY                                                                                                & 110th    & 0.9946                                                         & Energy                              \\
S.1108        & Algorithmic Accountability Act of 2019                                                                                & 116th    & 0.9938                                                         &                                     \\
H.R.1064      &                                                                                                                       & 116th    & 0.9929                                                         & Government Operations and Politics  \\
S.3484        & \begin{tabular}[c]{@{}l@{}}Grant Reporting Efficiency and Agreements \\ Transparency Act of 2018\end{tabular}         & 115th    & 0.9928                                                         & Government Operations and Politics  \\
H.R.7347      & \begin{tabular}[c]{@{}l@{}}Utilizing Significant Emissions with Innovative \\ Technologies Act\end{tabular}           & 115th    & 0.9928                                                         & Environmental Protection            \\
S.387         & Fair Chance to Compete for Jobs Act of 2019                                                                           & 116th    & 0.9927                                                         & Government Operations and Politics \\ 
\bottomrule
\end{tabular}
}
\end{table}

\begin{table}[]

\caption{The proportion of unlabelled bills with evidence of lobbying by subject area (top 10 highest probability subjects)}
\scalebox{0.75}{

\begin{tabular}{lcccc}
\toprule
subject                                     & \begin{tabular}[c]{@{}c@{}}Lobbied with \\ \textgreater{}50\% probability\end{tabular} & \begin{tabular}[c]{@{}c@{}}Lobbied with \\ \textgreater{}75\% probability\end{tabular} & \begin{tabular}[c]{@{}c@{}}Lobbied with \\ \textgreater{}90\% probability\end{tabular} & \begin{tabular}[c]{@{}c@{}}Total number of \\ unlabelled bills\end{tabular} \\
\midrule
Energy                                      & 0.4144                                                                                 & 0.1799                                                                                 & 0.0674                                                                                 & 1262                                                                  \\
Finance and Financial Sector                & 0.3737                                                                                 & 0.1698                                                                                 & 0.0554                                                                                 & 1678                                                                  \\
Commerce                                    & 0.3302                                                                                 & 0.1101                                                                                 & 0.0259                                                                                 & 1508                                                                  \\
Emergency Management                        & 0.3052                                                                                 & 0.1381                                                                                 & 0.0442                                                                                 & 724                                                                   \\
Science, Technology, Communications         & 0.2989                                                                                 & 0.1346                                                                                 & 0.0503                                                                                 & 1114                                                                  \\
Health                                      & 0.2850                                                                                 & 0.1221                                                                                 & 0.0420                                                                                 & 6453                                                                  \\
Labor and Employment                        & 0.2576                                                                                 & 0.0761                                                                                 & 0.0172                                                                                 & 1747                                                                  \\
Transportation and Public Works             & 0.2437                                                                                 & 0.0865                                                                                 & 0.0177                                                                                 & 2208                                                                  \\
Environmental Protection                    & 0.2309                                                                                 & 0.0939                                                                                 & 0.0271                                                                                 & 1884                                                                  \\
Immigration                                 & 0.2261                                                                                 & 0.0745                                                                                 & 0.0232                                                                                 & 1464                                                                  \\

\bottomrule
\end{tabular}
}
\label{tab:unlabelled_subject}

\end{table}

The above findings can imply two things. In the US all lobbying activity has to be reported but our findings suggest that there are bills that have not been filed under the Lobbying Disclosure Act (LDA), but carry the hallmarks of lobbied bills. First, these bills could have been indirectly affected by lobbying (i.e. were not lobbied but the legislator designed them in a way that made them similar to lobbied bills). There is also a possibility that not all bill-specific lobbying activity is reported, and the the OpenSecrets.org data is incomplete. 

Our proposed method could improve the data OpenSecrets.org holds, and could potentially contribute to the enforcement of the LDA by indicating the bills that were also likely to have been affected by lobbying but were not filed as such by the parties involved in the lobbying.

\section{Conclusion}\label{sec:conl}
\vspace{-4pt}
Many times the automation of handling large amounts of legal documents comes from the desire to improve work efficiency by substituting out human handling of cases. We believe our paper belongs in a different group. Even humans with the highest level of domain specific expertise on lobbying, legislation, and rent seeking would struggle to mark out those bills that had been targeted by lobbying. We propose the training of an algorithm to find patterns that distinguish lobbied bills from non-lobbied ones.

For the legal field to learn from this exercise, our future work will focus on a more detailed analysis of what factors are important in the distinction between the two types of bills. For this we would also like to perform more exhaustive experiments, looking at how our results change with time, with subject area, or with the identity of the lobbying organisation - all of which is available in our dataset - affects our results. Because our linear models perform well, it has the appeal to make it interpretable, which is important for social science applications. 

Moreover, we would also like to experiment with transfer learning and examine if our model, together with a small sample of labelled data from another English-speaking jurisdiction, can be used to predict lobbying activity in countries where such data is less easily available than in the US.

%\vspace{-3pt}
%\newpage
%\clearpage
%\bibliographystyle{IEEEtran}
\bibliographystyle{abbrv}
%\bibliography{disco,disco1,additional_refs}
\bibliography{disco}

\begin{thebibliography}{10}

\bibitem{aletras2016predicting}
N.~Aletras, D.~Tsarapatsanis, D.~Preo{\c{t}}iuc-Pietro, and V.~Lampos.
\newblock Predicting judicial decisions of the european court of human rights:
  A natural language processing perspective.
\newblock {\em PeerJ Computer Science}, 2:e93, 2016.

\bibitem{boella2011using}
G.~Boella, L.~Di~Caro, and L.~Humphreys.
\newblock Using classification to support legal knowledge engineers in the
  eunomos legal document management system.
\newblock In {\em Fifth international workshop on Juris-informatics (JURISIN)},
  2011.

\bibitem{RF-Breiman2001}
L.~Breiman.
\newblock Random forests.
\newblock {\em Machine Learning}, 45(1):5--32, Oct 2001.

\bibitem{brulle2018climate}
R.~J. Brulle.
\newblock The climate lobby: a sectoral analysis of lobbying spending on
  climate change in the usa, 2000 to 2016.
\newblock {\em Climatic change}, 149(3-4):289--303, 2018.

\bibitem{chalkidis2019deep}
I.~Chalkidis and D.~Kampas.
\newblock Deep learning in law: early adaptation and legal word embeddings
  trained on large corpora.
\newblock {\em Artificial Intelligence and Law}, 27(2):171--198, 2019.

\bibitem{dale2019law}
R.~Dale.
\newblock Law and word order: Nlp in legal tech.
\newblock {\em Natural Language Engineering}, 25(1):211--217, 2019.

\bibitem{de2014advancing}
J.~M. De~Figueiredo and B.~K. Richter.
\newblock Advancing the empirical research on lobbying.
\newblock {\em Annual review of political science}, 17:163--185, 2014.

\bibitem{farzindar2004legal}
A.~Farzindar and G.~Lapalme.
\newblock Legal text summarization by exploration of the thematic structure and
  argumentative roles.
\newblock In {\em Text Summarization Branches Out}, pages 27--34, 2004.

\bibitem{goldberg2017neural}
Y.~Goldberg.
\newblock Neural network methods for natural language processing.
\newblock {\em Synthesis Lectures on Human Language Technologies},
  10(1):1--309, 2017.

\bibitem{grasse2011influence}
N.~Grasse and B.~Heidbreder.
\newblock The influence of lobbying activityin state legislatures: Evidence
  from wisconsin.
\newblock {\em Legislative Studies Quarterly}, 36(4):567--589, 2011.

\bibitem{grossmann2013lobbying}
M.~Grossmann and K.~Pyle.
\newblock Lobbying and congressional bill advancement.
\newblock {\em Interest Groups \& Advocacy}, 2(1):91--111, 2013.

\bibitem{hachey2006extractive}
B.~Hachey and C.~Grover.
\newblock Extractive summarisation of legal texts.
\newblock {\em Artificial Intelligence and Law}, 14(4):305--345, 2006.

\bibitem{hill2013determinants}
M.~D. Hill, G.~W. Kelly, G.~B. Lockhart, and R.~A. Van~Ness.
\newblock Determinants and effects of corporate lobbying.
\newblock {\em Financial Management}, 42(4):931--957, 2013.

\bibitem{laband1988social}
D.~N. Laband and J.~P. Sophocleus.
\newblock The social cost of rent-seeking: First estimates.
\newblock {\em Public Choice}, 58(3):269--275, 1988.

\bibitem{li2018law}
P.~Li, F.~Zhao, Y.~Li, and Z.~Zhu.
\newblock Law text classification using semi-supervised convolutional neural
  networks.
\newblock In {\em 2018 Chinese Control and Decision Conference (CCDC)}, pages
  309--313. IEEE, 2018.

\bibitem{lopez1994rent}
R.~A. Lopez and E.~Pagoulatos.
\newblock Rent seeking and the welfare cost of trade barriers.
\newblock {\em Public Choice}, 79(1-2):149--160, 1994.

\bibitem{Pennington14glove}
J.~Pennington, R.~Socher, and C.~D. Manning.
\newblock Glove: Global vectors for word representation.
\newblock In {\em In EMNLP}, 2014.

\bibitem{sulea2017predicting}
O.-M. Sulea, M.~Zampieri, M.~Vela, and J.~Van~Genabith.
\newblock Predicting the law area and decisions of french supreme court cases.
\newblock {\em arXiv preprint arXiv:1708.01681}, 2017.

\bibitem{wongchaisuwat2017predicting}
P.~Wongchaisuwat, D.~Klabjan, and J.~O. McGinnis.
\newblock Predicting litigation likelihood and time to litigation for patents.
\newblock In {\em Proceedings of the 16th edition of the International
  Conference on Articial Intelligence and Law}, pages 257--260. ACM, 2017.

\bibitem{you2017ex}
H.~Y. You.
\newblock Ex post lobbying.
\newblock {\em The Journal of Politics}, 79(4):1162--1176, 2017.

\end{thebibliography}
\label{sec:References}

\newpage
\section{Appendix}\label{sec:extl}

\begin{figure}[ht!]
\vspace{-15pt}
\centering
	\subfloat[]{
		\includegraphics[width=1\columnwidth]{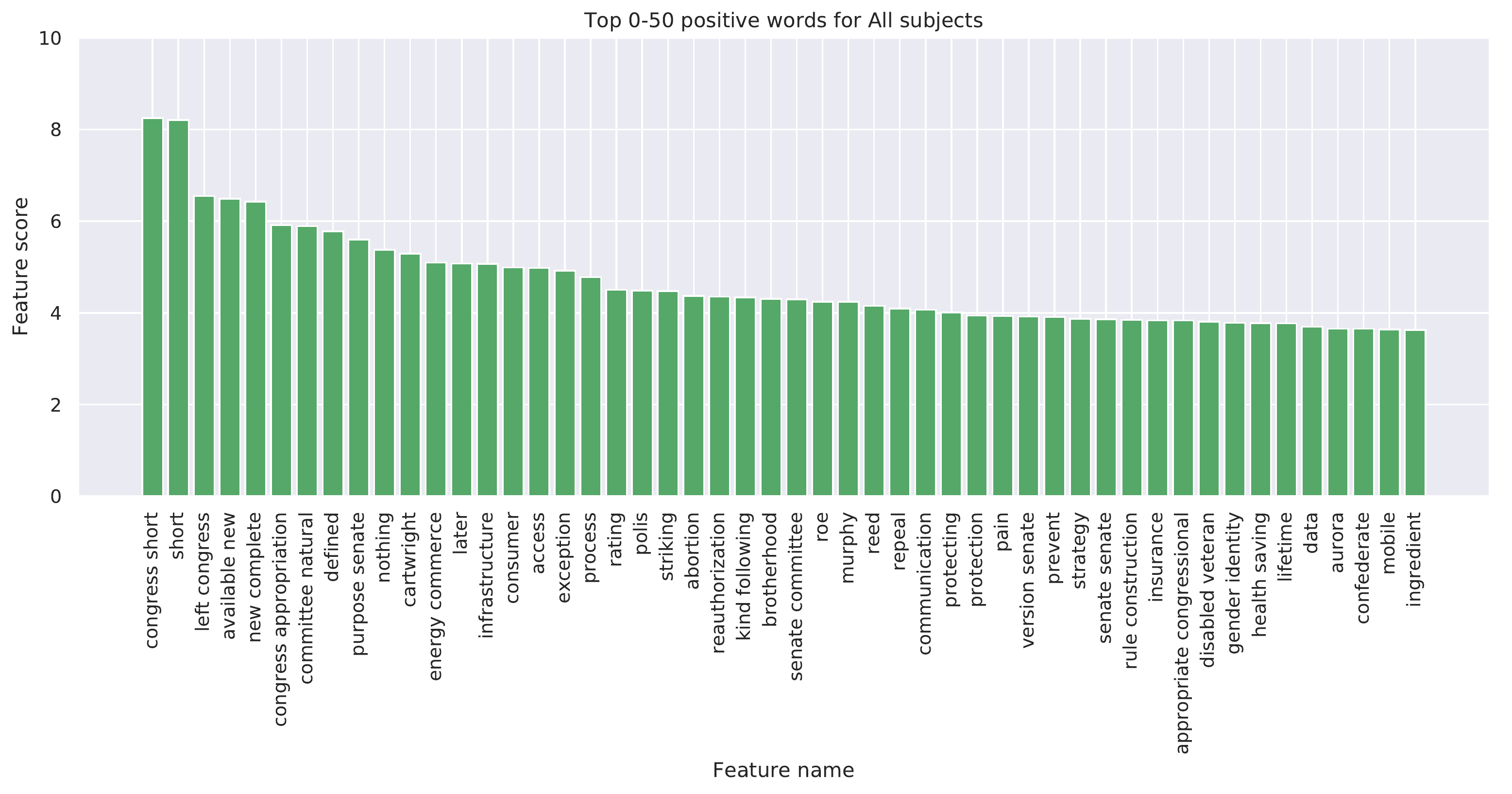}\label{fig:positive_words}
	}\\
	\subfloat[]{
		\includegraphics[width=1\columnwidth]{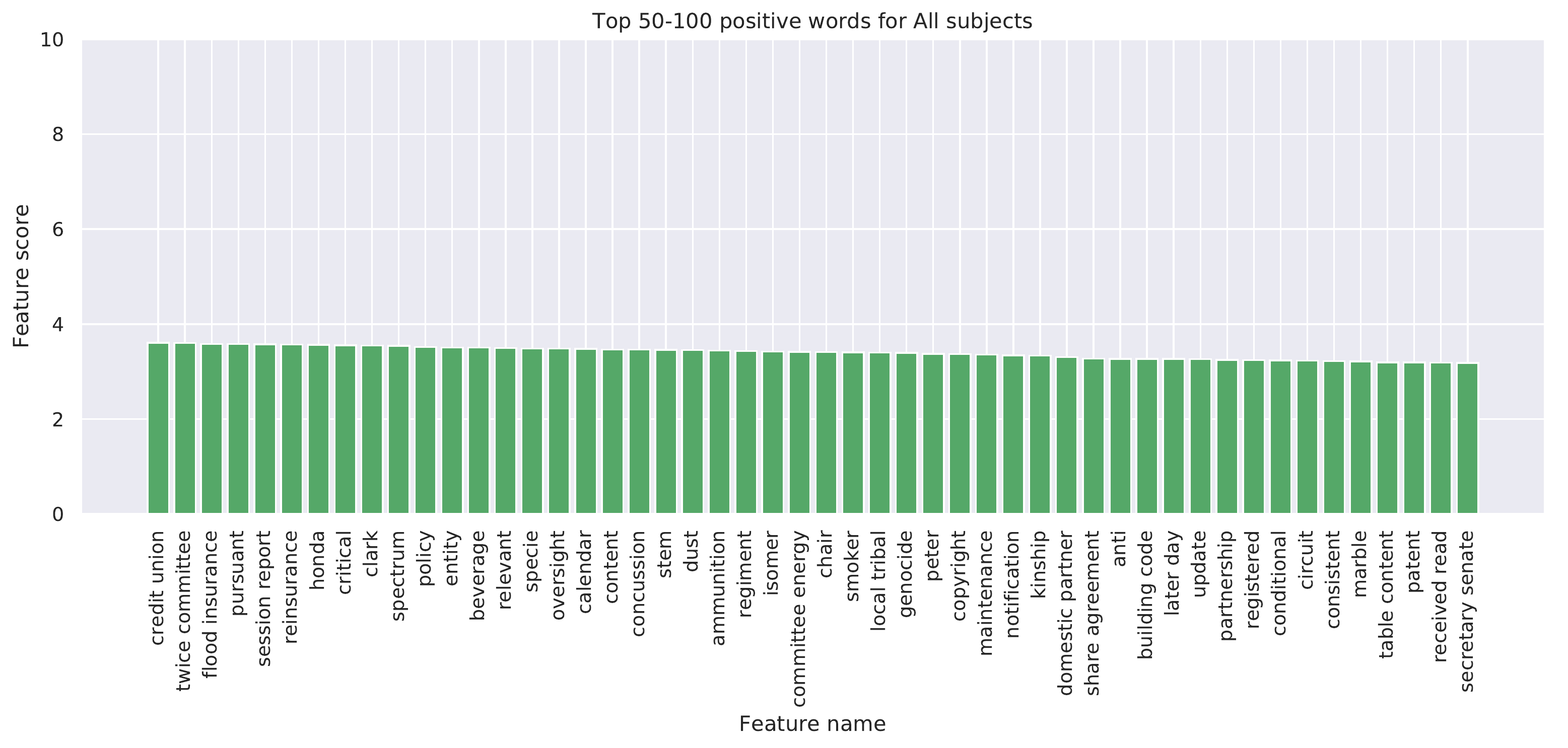}\label{fig:negative_words}
	}
\caption{Most common positive features across all subjects}	
\vspace{-15pt}
\end{figure}

\begin{figure}[ht!]
\vspace{-15pt}
\centering
	\subfloat[]{
		\includegraphics[width=1\columnwidth]{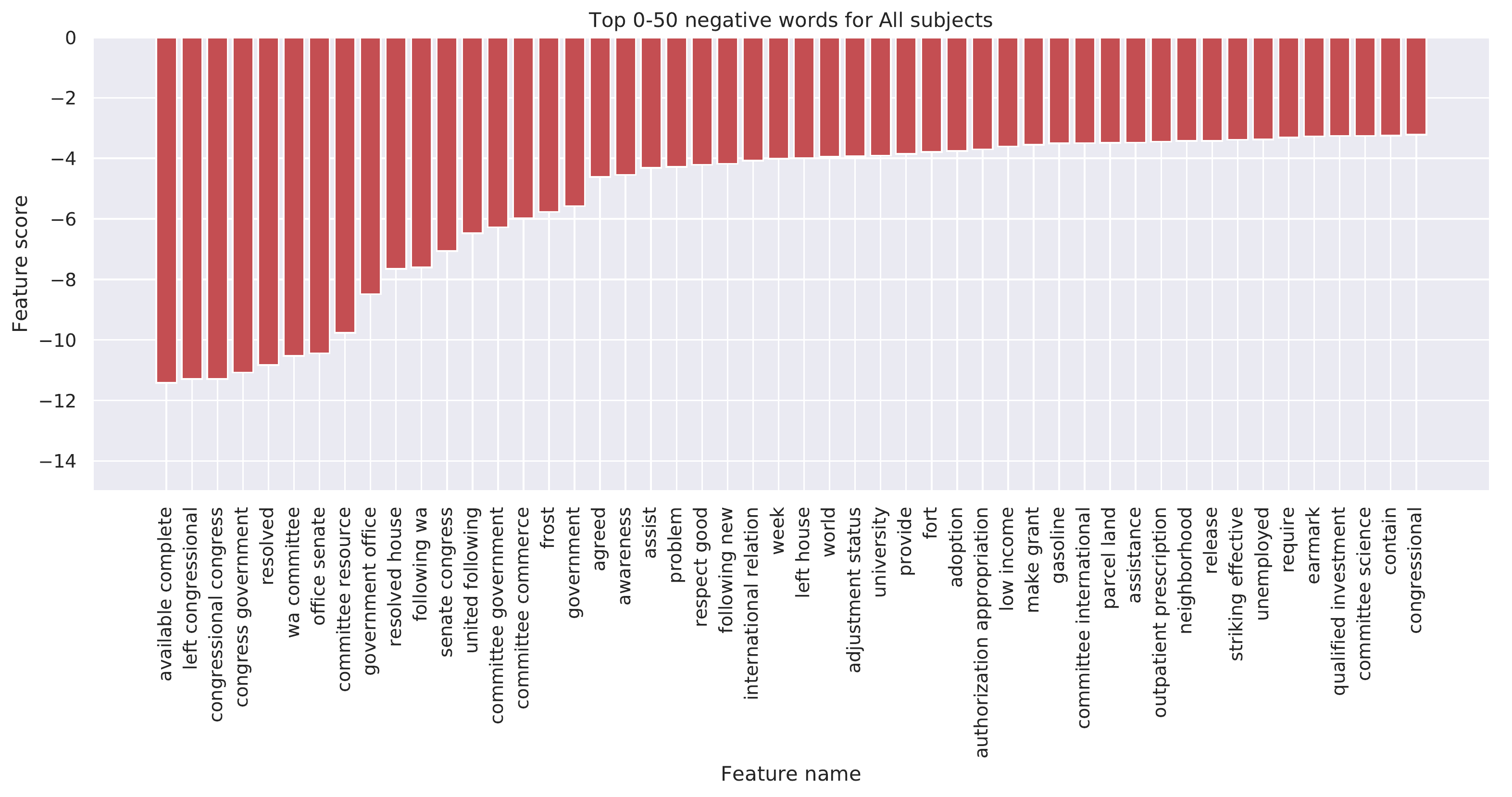}\label{fig:positive_words}
	}\\
	\subfloat[]{
		\includegraphics[width=1\columnwidth]{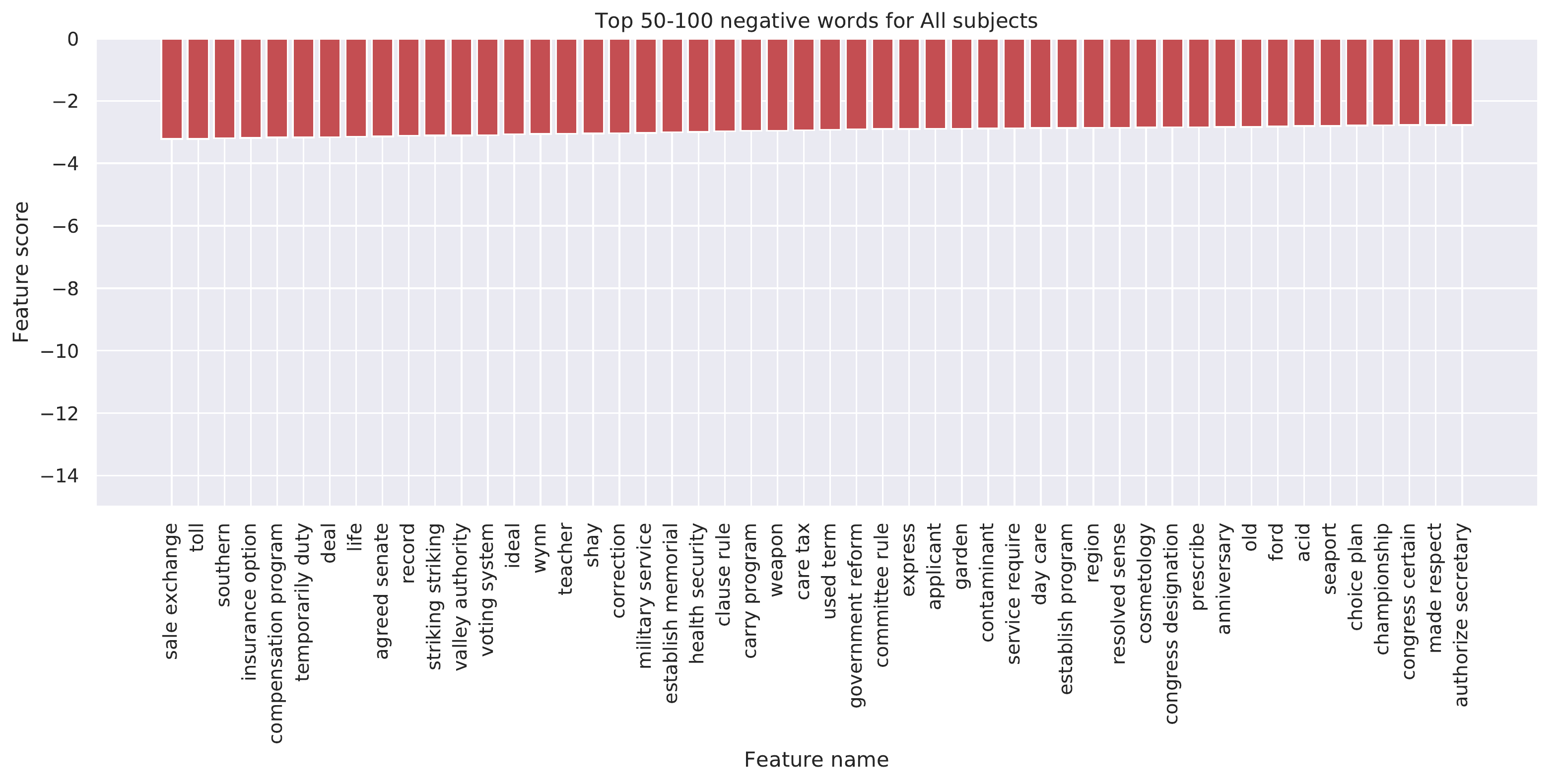}\label{fig:negative_words}
	}
\caption{Most common negative features across all subjects}	
\vspace{-15pt}
\end{figure}

\end{document}